\documentclass [superscriptaddress,floatfix,showpacs,preprint] {revtex4}

\usepackage {amsmath,amssymb,epsfig}
\begin {document}

\title {The influence of the rare earth ions radii on the
Low Spin to Intermediate Spin state transition in lanthanide
cobaltite perovskites: LaCoO$_3$ vs. HoCoO$_3$}

\author {I.A.~Nekrasov}
\affiliation {Institute of Metal Physics, Russian Academy of Sciences-Ural
Division, 620219 Yekaterinburg GSP-170, Russia}
\author {S.V.~Streltsov}
\affiliation {Institute of Metal Physics, Russian Academy of Sciences-Ural
Division, 620219 Yekaterinburg GSP-170, Russia}
\affiliation {Department of Theoretical Physics and Applied Mathematics,
Ural State Technical University, 620002 Yekaterinburg Mira 19, Russia}
\author {M.A.~Korotin}
\affiliation {Institute of Metal Physics, Russian Academy of Sciences-Ural
Division, 620219 Yekaterinburg GSP-170, Russia}
\author {V.I.~Anisimov}
\affiliation {Institute of Metal Physics, Russian Academy of Sciences-Ural
Division, 620219 Yekaterinburg GSP-170, Russia}

\date {\today}

\begin {abstract} We present first principles LDA+U calculations of electronic 
structure and magnetic state for LaCoO$_3$ and HoCoO$_3$. Low Spin to
Intermediate Spin state transition was found  in our calculations using
experimental crystallographic data for both materials with a much higher
transition temperature for HoCoO$_3$, which agrees well with the experimental
estimations. Low Spin state t$^6_{2g}$e$^0_g$ (non-magnetic) to Intermediate
Spin state t$^5_{2g}$e$^1_g$ (magnetic) transition of Co$^{3+}$ ions happens
due to the competition between crystal field t$_{2g}$-e$_g$ splitting and
effective exchange interaction  between 3$d$ spin-orbitals. We show that the
difference in crystal structure parameters for HoCoO$_3$ and LaCoO$_3$ due to
the smaller ionic radius of Ho ion comparing with La ion results in stronger
crystal field splitting for HoCoO$_3$ (0.09~eV $\approx$ 1000~K larger than for
LaCoO$_3$) and hence tip the balance between the Low Spin and Intermediate Spin
states to the non-magnetic solution in HoCoO$_3$. \end {abstract}

\pacs {78.70.Dm, 71.25.Tn}

\maketitle

\section {Introduction}

In lanthanum cobaltite LaCoO$_3$ there is a temperature induced
transition from a non-magnetic to a magnetic
state, which has attracted considerable interest last years \cite {Imada98}. In
the ground state of LaCoO$_3$ there is no magnetic moment on Co ions. At low
temperature the magnetic susceptibility increases exponentially
with temperature, exhibiting a maximum near 100~K. While initially
this maximum was ascribed to a transition from a Low Spin
non-magnetic ground state (t$^6_{2g}$,~S=0) to a High Spin
state (t$^4_{2g}$e$_g^2$,~S=2), later a new scenario involving
Intermediate Spin state (t$^5_{2g}$e$_g^1$,~S=1) has been proposed.
Using the results of LDA+U, Korotin {\it et al.} \cite {Korotin96} explained
stabilization of Intermediate Spin state over High Spin state
due to the large hybridization between
Co-3$d$~(e$_{g}$) and O-2$p$ orbitals.

\begin {figure}[htb]
\centering
\epsfig {file=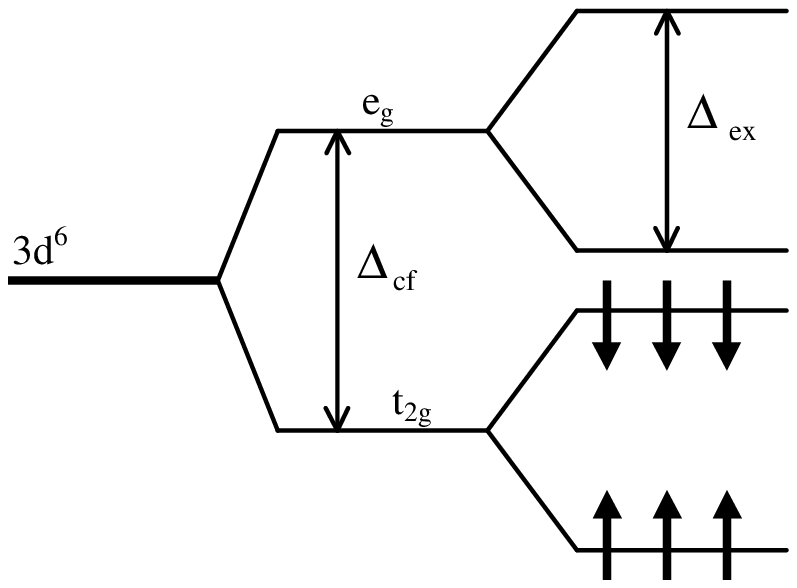,width=0.4 \textwidth}
\caption {Scheme of energy levels for Low Spin state of Co$^{3+}$ ions in
LaCoO$_3$.}
\label {goodenough}
\end {figure}

The physical reason for this transition is a competition between
crystal-field energy $\Delta_{cf}$
(t$_{2g}$-e$_g$ energy splitting) and intra-atomic (Hund) exchange
energy $\Delta_{ex}$ (Fig.~\ref  {goodenough}). In the ground state
$\Delta_{cf}$ is only slightly larger then $\Delta_{ex}$ so that the energy of
the excited magnetic state of Co ion is relatively small and with increasing of
temperature its population increases resulting in an increase of magnetic
susceptibility.
When La ions are substituted in perovskite cobaltites by other
rare earth elements with
smaller ionic radii, magnetic properties show significant changes.
For NdCoO$_3$ material $^{59}$Co Knight shift measurements \cite {itoh}
showed that Co ions remain to be in the Low Spin state up to 580~K,
where the gradual metal-insulator transition is observed.
High-temperature diffraction study of lanthanide cobaltite perovskites
LnCoO$_3$ (Ln=Nd, Gd, Dy and Ho) \cite {liu}
led its authors to conclusion, that ``at room temperature all cobalt ions are
in the Low Spin state regardless of the Ln atomic number'' and only as
temperature increases to 1000 K ``the possible electronic phase transition
can be suggested''. Magnetic and electric measurements for
Ho$_{1-x}$Ca$_x$CoO$_3$ system \cite {im} showed that ``there is no apparent
transition between Low Spin and High Spin of Co$^{3+}$ ions in 300-900~K
temperature range''.
All these data show that Low Spin state of Co$^{3+}$ ions in
lanthanide cobaltite perovskites with smaller rare earth ions
becomes more stable comparing with LaCoO$_3$ and transition temperature
increases dramatically. In the present work we have investigated relative
stability of Low Spin and Intermediate Spin states of Co$^{3+}$ ions
in LaCoO$_3$ and HoCoO$_3$ as a function of experimental crystal structure
parameters measured at different temperatures. We have found that, indeed,
while for LaCoO$_3$ transition temperature value (defined as the temperature,
where  calculated total energy of  Low Spin solution becomes higher than
the energy of Intermediate Spin solution) is 140~K, for HoCoO$_3$ this value
is 1070~K. The physical reason for this is decreasing of Co-O bond length
due to the chemical pressure happening with substitution of La ions by smaller
Ho ions and hence increasing of the crystal field t$_{2g}$-e$_g$
energy splitting value thus stabilizing Low Spin state. Such conclusion is
supported by the results of the recent diffraction experiments under pressure
for LaCoO$_3$ \cite {vogt}, which were interpreted as ``pressure-induced
Intermediate-to-Low Spin-state transition''.

\section {Crystal structure }
\label {crystal}

\begin {figure}[htb]
\centering
\epsfig {file=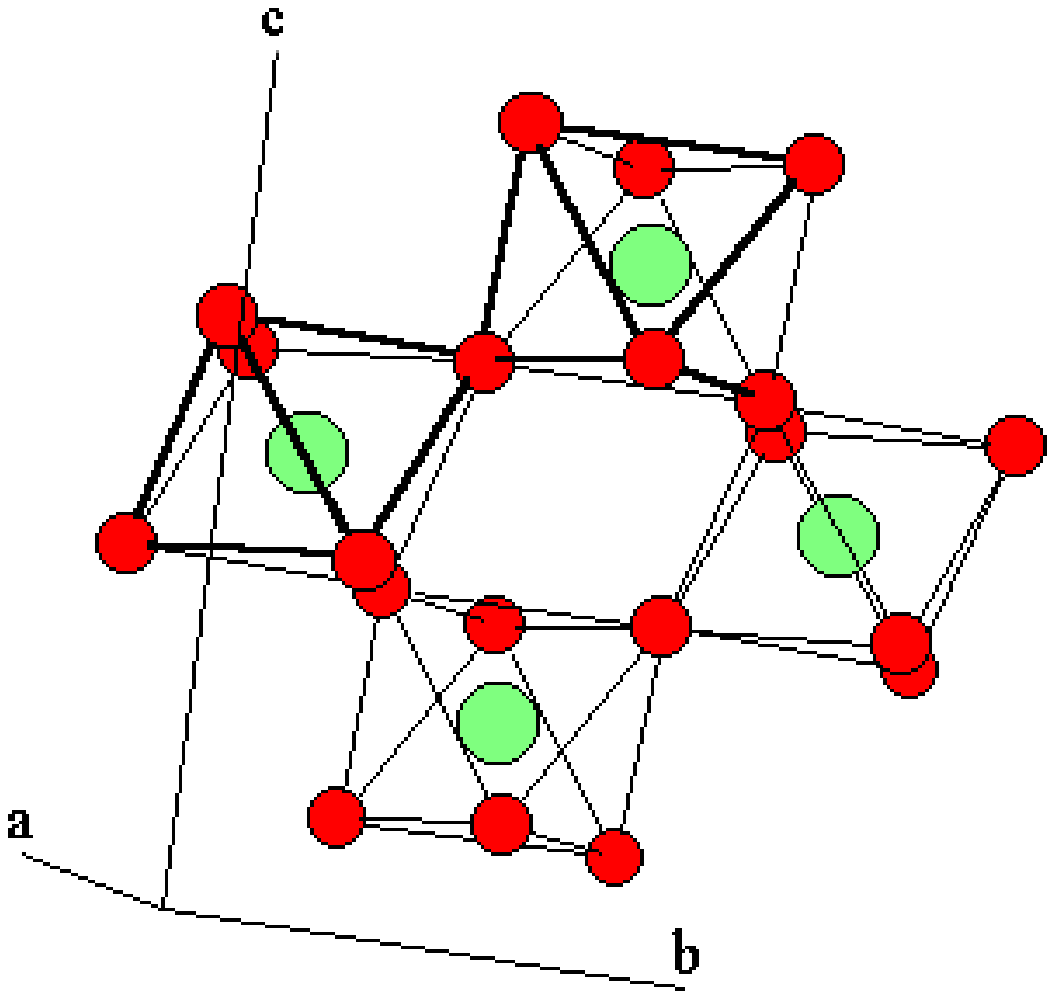,height=0.24\textheight}
\rule{0.05\textwidth}{0pt}
\epsfig {file=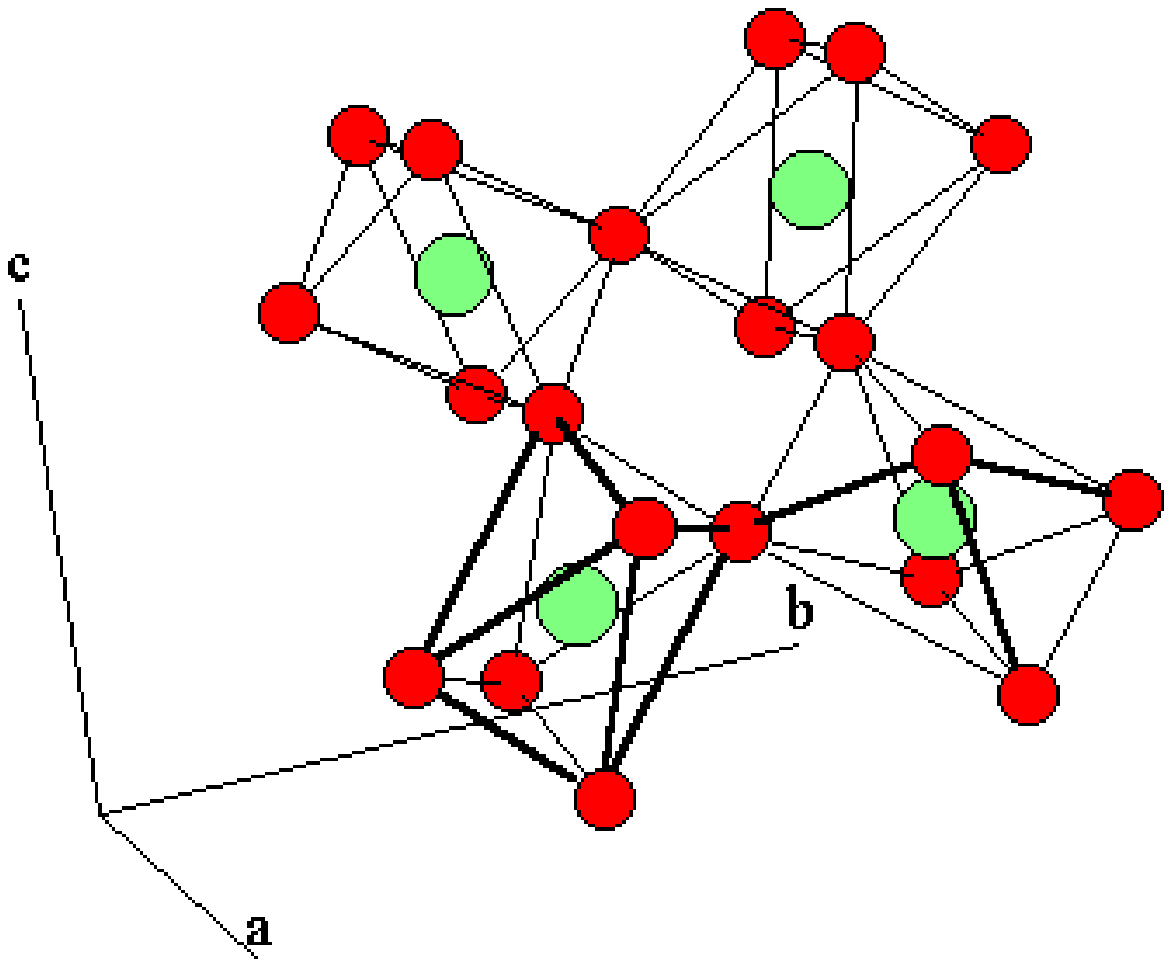,height=0.24\textheight}
\caption {The rhombohedral crystal structure of LaCoO$_3$ (left) and
the orthorhombic crystal structure of HoCoO$_3$ (right). Co: large
spheres; O: small spheres.}
\label {lahoco_cs}
\end {figure}

LaCoO$_3$ crystallizes in the rhombohedrally distorted cubic
perovskite structure \cite {Thornton86,Radaelli02}. According to the latest
crystallographic data by Radaelli {\it{et al.}} \cite {Radaelli02} the
space group of this compound is R$\bar 3$c for all measured
temperatures and has two formula units in the unit cell.  Main
structural motif of this compound is a nearly perfect CoO$_6$ octahedron.
The rhombohedral distortion of the parent cubic perovskite structure
can be described by deformation along the body diagonal so that
the angle of the Co-O-Co bond has changed form 180$^\circ$  to
$\sim$163$^\circ$ (Fig. \ref {lahoco_cs}).

HoCoO$_3$ has orthorhombically distorted perovskite crystal structure (prototype
GdFeO$_3$, space group $Pbnm$) \cite {liu} (Fig. \ref {lahoco_cs}), which
contains four formula units in the unit cell. Like for  LaCoO$_3$ the main
crystallographic motif form CoO$_6$ octahedra, which are rotated in $ab$-plane
and tilted in respect to $c$-axis. Due to smaller ionic radius of
Ho ion comparing with La ion, distortion from the cubic perovskite structure
is stronger for HoCoO$_3$  than for LaCoO$_3$ and the angles of the Co-O-Co bond
have values 149$^\circ$ (in $ab$-plane) and 152$^\circ$ (along $c$-axis).

In addition to the strong bending of the Co-O-Co bond the average length of the
Co-O bond is decreased with substitution of La by Ho (1.934~\AA \ and
1.921~\AA \ at 300~K \cite {Radaelli02,liu}).
Such decrease can be understood as a result of chemical pressure, induced by
the smaller ionic radius of Ho ion comparing with La ion.

The crystal structure parameters as a function of temperature were measured for
LaCoO$_3$ in \cite {Radaelli02} and for HoCoO$_3$ in \cite {liu}. It was
demonstrated in \cite {Korotin96} that LDA+U calculations can reproduce a Low Spin-Intermediate Spin
transition for LaCoO$_3$ using temperature dependent crystal structure as an
input. We used the same procedure for HoCoO$_3$ in order to find if
a small decrease of Co-O bond length due to the chemical pressure
can results in stabilization of the Low Spin state.

Five 3d-orbitals are separated into triply degenerate t$_{2g}$ and
doubly degenerate e$_g$ subsets only in cubic symmetry lattice.
The rhombohedral crystal structure of LaCoO$_3$ and 
the orthorhombic crystal structure of HoCoO$_3$ have lower symmetry than cubic
as a whole but CoO$_6$ octahedra are only slightly distorted in 
those compounds. In local coordinate system centered
on Co ion with the axes directed to the oxygen ions t$_{2g}$-e$_g$  
orbitals are still well defined. In the following
we have used such defined orbitals for analisys of the calculated
electronic structure.

\section {LDA band structure}
\label {lda_bands}

The standard L(S)DA (Local (Spin) Density Approximation \cite {kohn65,hedin})
based methods can not describe magnetic state for LaCoO$_3$ without taking into
account Coulomb interaction between 3$d$ electrons as it was demonstrated
in \cite {Korotin96}. However analysis of the LDA results is instructive
for understanding of the basic electronic structure of
lanthanide cobaltite perovskites.

\begin {figure}[htb]
\vspace{4mm}
\centering
\epsfig {file=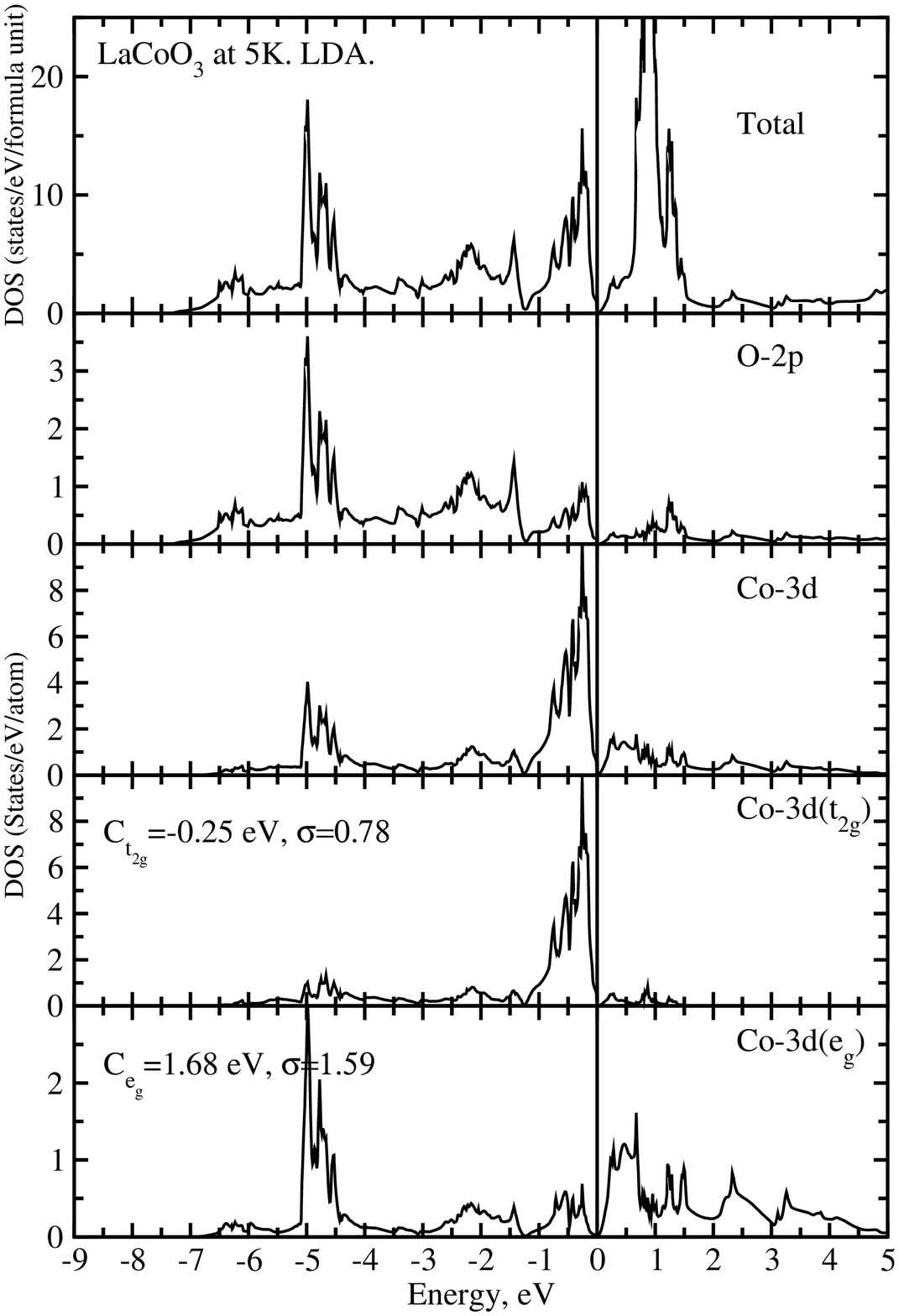,  width=0.4\textwidth}
\rule{0.05\textwidth}{0pt}
\epsfig {file=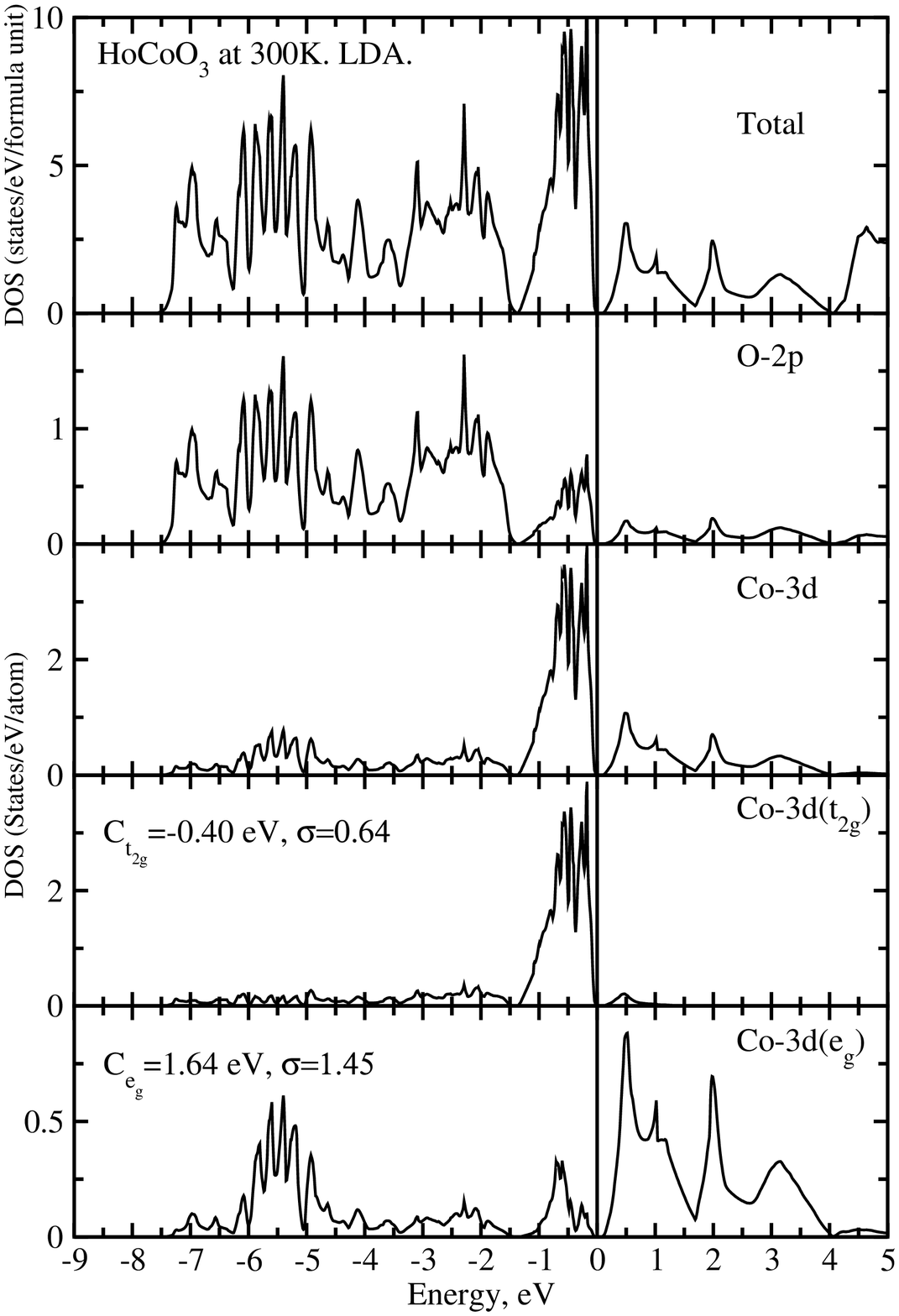,width=0.4\textwidth}
\caption {DOS of LaCoO$_3$ (left) and HoCoO$_3$ (right) calculated within the
LDA. Top panel: total DOS; second from top panel~- partial O-2$p$ DOS; Last
three panels: partial Co-3$d$, Co-3$d$~(t$_{2g}$) and Co-3$d$~(e$_{g}$) DOS.
For partial Co-3$d$~(t$_{2g}$) and Co-3$d$~(e$_{g}$) DOS
centers of gravities C and standard deviation $\sigma$
of corresponding Co 3$d$ orbitals are shown.
The Fermi level is zero energy.}
\label {lahoco_lda}
\end {figure}

We used LMTO method \cite {LMTO} to
perform calculations of  electronic structure for
LaCoO$_3$ and HoCoO$_3$. The results are presented in
Fig.~\ref {lahoco_lda}.  The radii of the muffin-tin
spheres in LaCoO$_3$ for 5~K structure were  R$_{\rm La }$=3.76~a.u.,
R$_{\rm Co}$=2.49~a.u., and R$_{\rm O}$=2.0~a.u.  With temperature
increasing R$_{\rm Co}$ and R$_{\rm O}$ were increased proportionally
to Co-O distance for a given temperature. In HoCoO$_3$ for lowest
temperature (300~K) crystal structure R$_{\rm Co}$ and R$_{\rm O}$ were
taken as the corresponding values for LaCoO$_3$  scaled
in proportion of Co-O distances in LaCoO$_3$ and HoCoO$_3$.
Then R$_{\rm Co}$ and R$_{\rm O}$ were increased
proportionally to the change of the Co-O distance in HoCoO$_3$ for a given
temperature. R$_{\rm La }$  was adjusted to fill completely the
volume of the unit cell.
In the orbital basis set the following states were included:
Co(4$s$,4$p$,3$d$), O(3$s$,2$p$), La,Ho(6$s$,6$p$,5$d$,4$f$). Partially
filled 4$f$ states of Ho were treated as a pseudo-core states.

Figure \ref {lahoco_lda} (left panel) shows in details LDA calculated electronic
band structure of LaCoO$_3$. On the top of
the figure  total density of states (DOS) is shown. There are three
distinguishable sets of bands: completely filled O-$p$ bands,
partially filled Co-$d$ bands and empty La-$f$ bands.  The bands in the
energy range from -7~eV to -1.5~eV originate mainly from O-2$p$ states
but have a significant admixture from Co-3$d$ states. Between 0~eV and 2~eV
there are empty La-4$f$ bands. The two groups of bands presented in the third
panel, which extend from -1.5~eV  to~0 eV and from 0~eV to 2.0~eV, are the
Co-3$d$ states. Also a substantial O-2$p$ contributions are apparent in
this energy range. Co-3$d$ states, due to the octahedral symmetry as we
discussed above, are formed by completely occupied t$_{2g}$ states and
empty e$_g$ states with a ``pseudo-gap'' between them. The partial densities of
states for t$_{2g}$ and e$_g$ states  are
presented in the fourth and fifth panels of Fig.~\ref {lahoco_lda},
correspondingly. We have calculated centers of gravity and standard deviations
of partial t$_{2g}$ and e$_g$ DOS (shown in Fig.~\ref {lahoco_lda}) in order to
determine the value of crystal field splitting between t$_{2g}$ and e$_g$
states.  It was found that
t$_{2g}$ states are  1.93~eV lower than e$_g$. Also
t$_{2g}$ band is much more narrow than e$_g$.

LDA band structure obtained for HoCoO$_3$
is shown in Fig.~\ref {lahoco_lda} (right panel).
In general it is very similar to LaCoO$_3$ bands (except for the absence
of 4$f$ band because Ho-4$f$ states were treated as a pseudo-core).
The  important
difference  from LaCoO$_3$ is relative positions of t$_{2g}$ and e$_g$
bands of Co-3$d$ shell. As one can see from  last two panels of
Fig.~\ref {lahoco_lda} (right side), t$_{2g}$ states are  2.04~eV lower than e$_g$
comparing with 1.93~eV for LaCoO$_3$
and also e$_g$ band is more narrow (standard deviation $\sigma$ is
1.45 comparing with 1.59 for LaCoO$_3$). The higher position of  e$_g$
states and the smaller width of the corresponding band result
in  opening of a small energy gap $\approx$0.07 eV.

Those results can be well understood from the difference in crystal structure of
two compounds.
The values of Co-O-Co bond angles deviate from 180$^\circ$ much more
in HoCoO$_3$ than in  LaCoO$_3$. That leads to a weaker
$d$-$p$-$d$ hybridization and hence to a more narrow  Co-3$d$~(e$_g$) band.

In adition to  Co-O-Co bond bending, substitution of La ion on a smaller
Ho ion lead to the compression of Co-O bond lengths. 
Decreased value of Co-O bond length in HoCoO$_3$ comparing with
LaCoO$_3$ results in the increased value of hybridization strength
between O-2$p$ states and Co-3$d$ states and hence to the increasing of
t$_{2g}$-e$_g$ energy splitting of Co-3$d$ states.

For the problem of Low Spin to Intermediate Spin state transition
the most important is increasing of the t$_{2g}$-e$_g$ energy
splitting value. As magnetic transition in cobaltites is determined by the competition
between crystal field and exchange energies, this result can dramatically
increase excitation energy to the magnetic state and hence increase
transition temperature for HoCoO$_3$ comparing with  LaCoO$_3$.

One can estimate the change of excitation to magnetic state energy
as equal to the change in the value of  t$_{2g}$-e$_g$ energy splitting:
2.04-1.93=0.09~eV$\approx$1000K. This value must be connected to the
transition temperature difference. We will show below, that this crude estimate
agrees surprisingly well with the results of our LDA+U calculations.

\section {LDA+U results}
\label {transit}

In the present work we used a general rotationally invariant formulation
of LDA+U approach instead of the old version used in \cite{Korotin96}.
The main idea of the LDA+U approach \cite {ldau1,ldau2} is to add to the LSDA
functional the term $E^U$ corresponding to the mean-field approximation of the
Coulomb interaction in multiband Hubbard model (Hartree-Fock approximation)

\begin {equation}  \label {U1}
E^{LDA+U}[\rho ^\sigma ({\bf r}),\{n^\sigma \}]=E^{LSDA}[\rho ^\sigma ({\bf %
r)}]+E^U[\{n^\sigma \}]-E_{dc}[\{n^\sigma \}],
\end {equation}
$\rho ^\sigma ({\bf r})$ is the charge density for spin-$\sigma $
electrons, $E^{LSDA}[\rho ^\sigma ({\bf r})]$ is the standard LSDA
(Local Spin-Density Approximation) functional and Coulomb interaction
term $E^U$ is a functional of the occupation matrix $n^\sigma$:
\begin {equation}
\label{Occ}n_{mm^{\prime }}^\sigma =-\frac 1\pi
\int^{E_F}ImG_{inlm,inlm^{\prime }}^\sigma (E)dE.
\end {equation}
Here $G_{inlm,inlm^{^{\prime }}}^\sigma (E)=\langle inlm\sigma \mid
(E-\widehat{H})^{-1}\mid inlm^{^{\prime }}\sigma \rangle $ are the
elements of the Green function matrix in the basis of $d$-orbitals
$\mid inlm\sigma \rangle $ ($i$ denotes the cite, $n$ the main quantum
number, $l$- orbital quantum number, $m$- magnetic number and $\sigma
$- spin index)
\begin {equation}  \label {upart}
\begin {array}{c}
E^U[\{n^\sigma\}]=\frac 12\sum_{\{m\},\sigma }\{\langle m,m^{\prime \prime }\mid
V_{ee}\mid m^{\prime },m^{\prime \prime \prime }\rangle n_{mm^{\prime
}}^\sigma n_{m^{\prime \prime }m^{\prime \prime \prime }}^{-\sigma }+ \\\\
(\langle m,m^{\prime \prime }\mid V_{ee}\mid m^{\prime },m^{\prime \prime
\prime }\rangle -\langle m,m^{\prime \prime }\mid V_{ee}\mid m^{\prime
\prime \prime },m^{\prime }\rangle )n_{mm^{\prime }}^\sigma n_{m^{\prime
\prime }m^{\prime \prime \prime }}^\sigma \},
\end {array}
\end {equation}
where $V_{ee}$ are the screened Coulomb interactions among the
\textit{d} electrons. Finally, the last term in Eq.~(\ref {U1}) corrects
for double counting and is given by

\begin {equation}  \label {U3}
E_{dc}[\{n^\sigma \}]=\frac 12UN(N-1)-\frac 12J[N^{\uparrow }(N^{\uparrow
}-1)+N^{\downarrow }(N^{\downarrow }-1)],
\end {equation}
with $N^\sigma =Tr(n_{mm^{\prime }}^\sigma )$ and $N=N^{\uparrow
}+N^{\downarrow }$. $U$~and~$J$ are screened Coulomb and exchange
parameters~\cite {Gunnarsson89,ucalc}.

In addition to the usual LSDA potential, an effective single-particle
potential to be used in the effective single-particle Hamiltonian has
the form:
\begin {equation}  \label{hamilt}
\widehat{H}=\widehat{H}_{LSDA}+\sum_{mm^{\prime }}\mid inlm\sigma \rangle
V_{mm^{\prime }}^\sigma \langle inlm^{\prime }\sigma \mid,
\end {equation}
\begin {equation}  \label {Pot}
\begin {array}{c}
V_{mm^{\prime }}^\sigma =\sum_{m^{\prime \prime}m^{\prime \prime \prime}}
\{\langle m,m^{\prime \prime }\mid
V_{ee}\mid m^{\prime },m^{\prime \prime \prime }\rangle n_{m^{\prime \prime
}m^{\prime \prime \prime }}^{-\sigma }+ \\
\\
(\langle m,m^{\prime \prime }\mid V_{ee}\mid m^{\prime },m^{\prime \prime
\prime }\rangle -\langle m,m^{\prime \prime }\mid V_{ee}\mid m^{\prime
\prime \prime },m^{\prime }\rangle )n_{m^{\prime \prime }m^{\prime \prime
\prime }}^\sigma \}- \\
\\
U(N-\frac 12)+J(N^{\sigma}-\frac 12).
\end {array}
\end {equation}

The matrix elements of Coulomb interaction can be expressed in terms of complex
spherical harmonics and effective Slater integrals $F^k$~\cite {JUDD} as
\begin {equation}  \label {slater}
\langle m,m^{\prime \prime }\mid V_{ee}\mid m^{\prime },m^{\prime \prime
\prime }\rangle =\sum_{k=0}^{2l} a_k(m,m^{\prime },m^{\prime \prime },m^{\prime \prime
\prime })F^k,
\end {equation}
and
\[ a_k(m,m^{\prime },m^{\prime \prime },m^{\prime \prime \prime
})=\frac{4\pi}{2k+1}\sum_{q=-k}^k\langle lm\mid Y_{kq}\mid lm^{\prime }\rangle
\langle lm^{\prime \prime}\mid Y_{kq}^{*}\mid lm^{\prime \prime \prime }\rangle.
\] For $d$ electrons one needs $F^0, F^2$ and $F^4$ and these can
be linked to the Coulomb and Stoner parameters $U$ and $J$ obtained
from the LSDA-supercell procedures via $U=F^0$ and $J=(F^2+F^4)/14$,
while the ratio $F^2/F^4$ is to a good accuracy a constant $\sim 0.625$
for the 3$d$ elements \cite {deGroot,ANISOL}
(for $f$ electrons the corresponding expression
is $J=(286F^2+195F^4+250F^6)/6435$). The Coulomb parameter $U$ is
calculated as a second derivative of the total energy (or the first
derivative of the corresponding eigenvalue)  in respect to the
occupancy of localized orbitals of the central atom in a supercell with
fixed occupancies on all other atoms \cite {Gunnarsson89}.

\begin {figure}[htb]
\vspace{3mm}
\centering
\epsfig {file=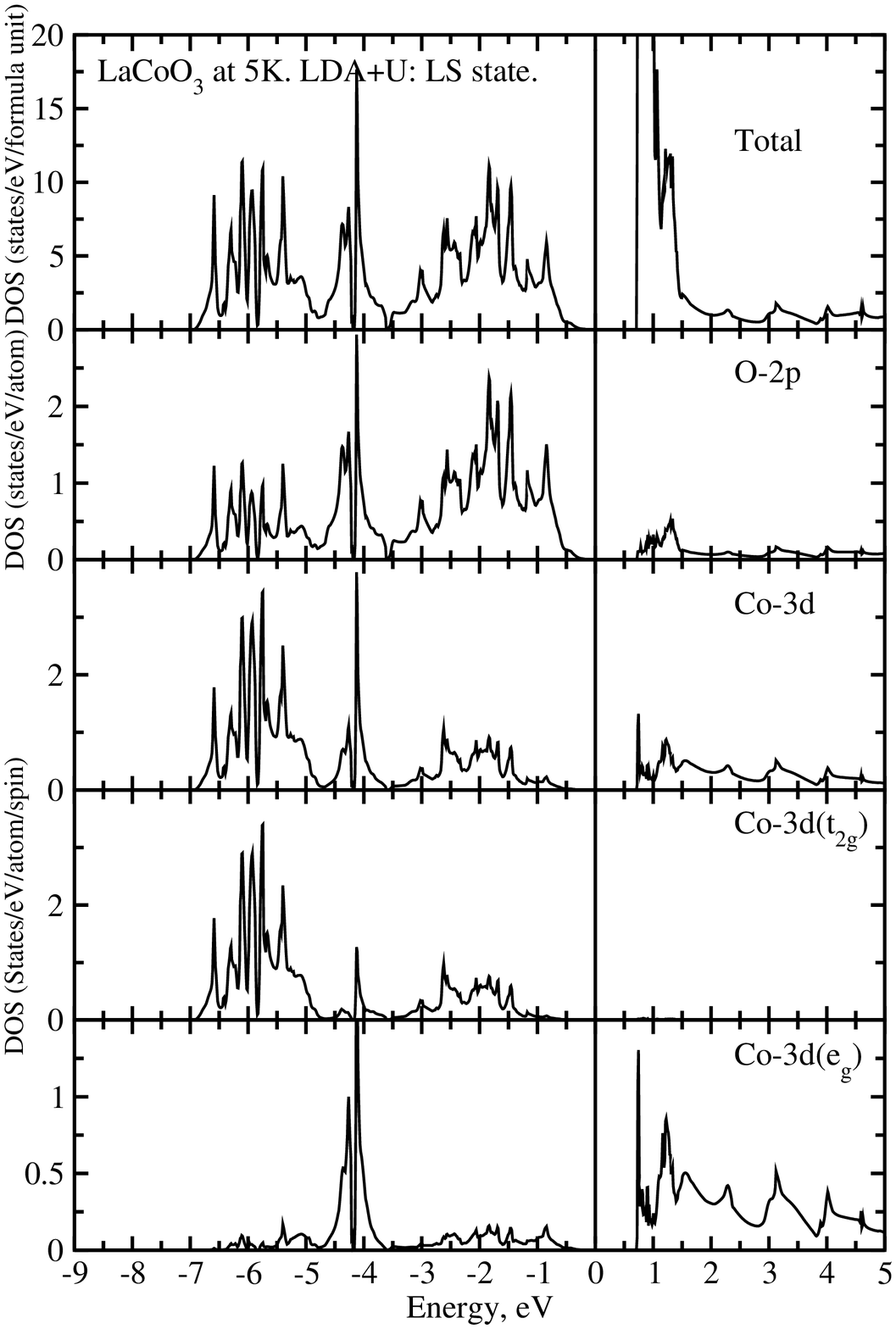,  width=0.4\textwidth}
\rule{0.05\textwidth}{0pt}
\epsfig {file=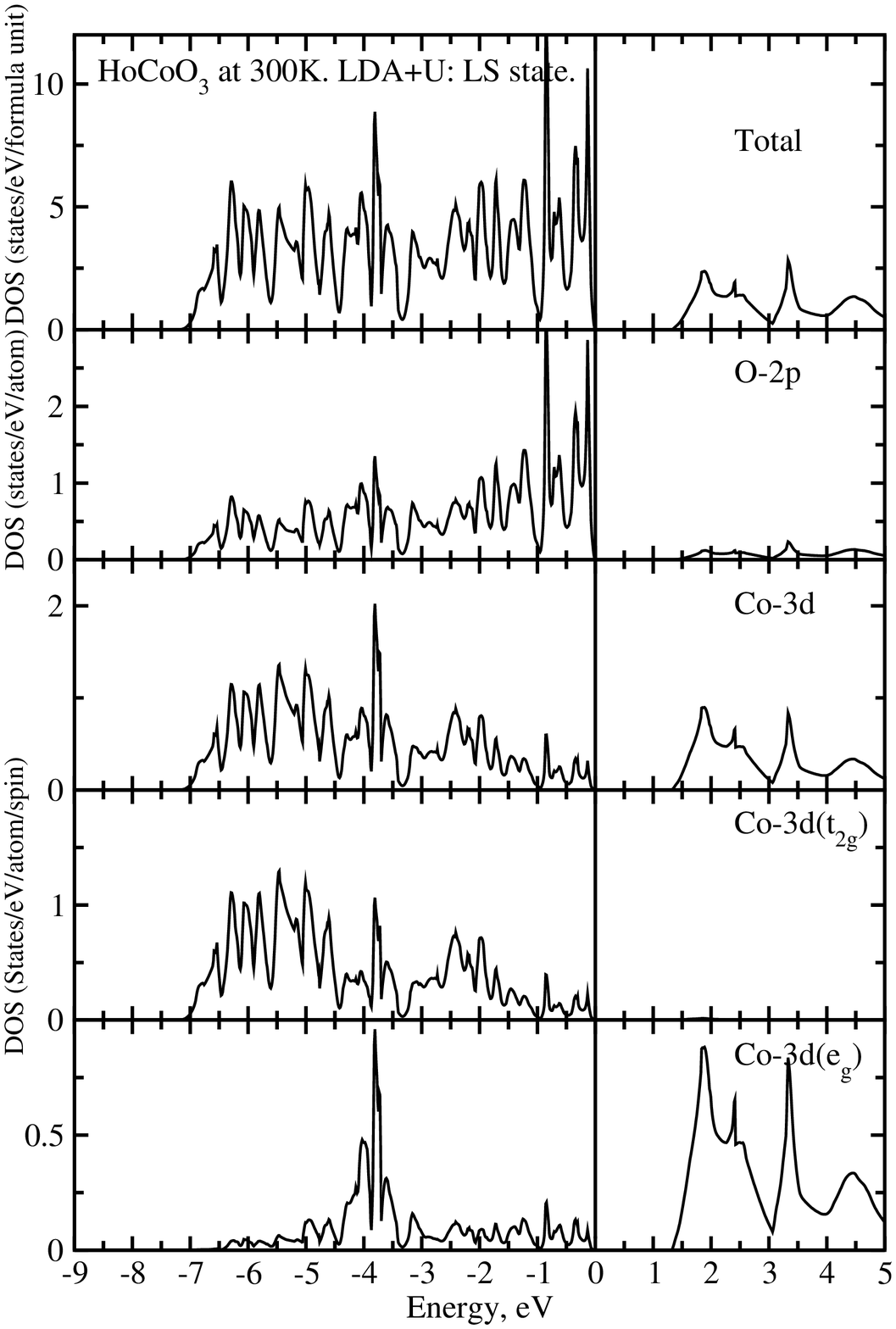,width=0.4\textwidth}
\caption {DOS of LaCoO$_3$ (left) and HoCoO$_3$ (right) calculated within the LDA+U
approach for Low Spin state t$^6_{2g}$e$^0_g$ of Co ion. Top panel:
total DOS; second from top panel: partial O-2$p$ DOS; Last three panels: partial
Co-3$d$, Co-3$d$~(t$_{2g}$) and Co-3$d$~(e$_g$) DOS correspondingly.
The Fermi level is zero energy.}
\label {lahoco_LS}
\end {figure}

Coulomb interaction parameters $U$=7.80~eV and $J$=0.99~eV used in our
LDA+U calculations were computed in \cite{Korotin96}
by the constrained LDA approach \cite{Gunnarsson89}.
The main effect of LDA+U potential correction (\ref{Pot})
is the energy splitting between occupied and empty states
in such a way that the former are pushed down and the latter up
comparing with LDA. In the result magnetic state solution, which does not
exist in LSDA, becomes stable in LDA+U and by comparing the relative energy
of two solutions, non-magnetic and magnetic, one can find the ground state
of the system.

We have investigated two solutions for both compounds for all crystal
structure parameters corresponding to the temperatures 5-300~K
for LaCoO$_3$ and 300-1080~K for HoCoO$_3$. The Low Spin
state solutions (Fig.~\ref{lahoco_LS})
are very similar to LDA results (Fig.~\ref{lahoco_lda}).
The main effect of LDA+U correction is higher position of the empty
Co-3$d$~(e$_g$) states and the opposite effect on the occupied
Co-3$d$~(t$_{2g}$) states and hence opening of a sizable energy gap
(0.7~eV for LaCoO$_3$ and 1.3~eV for HoCoO$_3$).

\begin {figure}[htb]
\vspace{3mm}
\centering
\epsfig {file=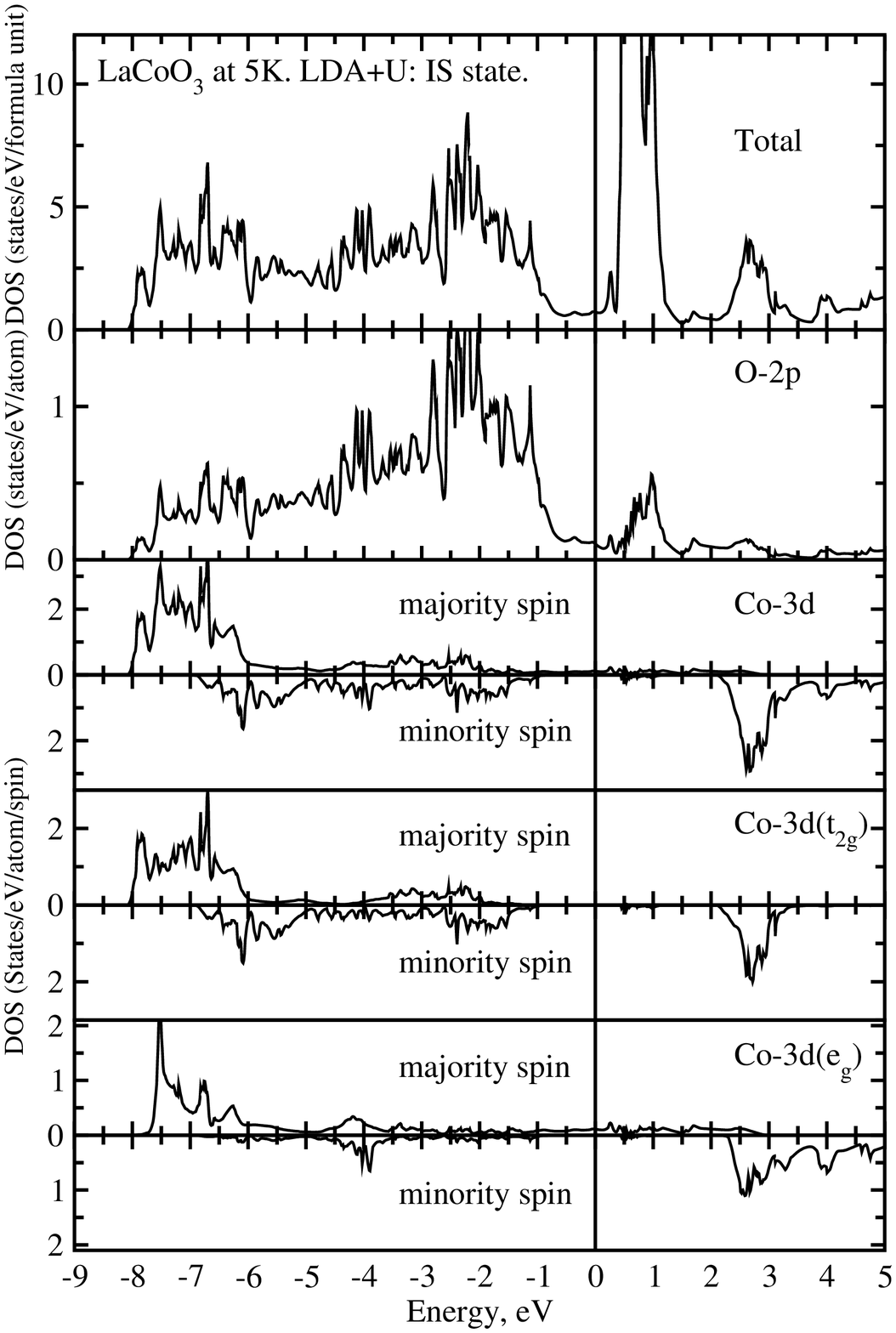,  width=0.4\textwidth}
\rule{0.05\textwidth}{0pt}
\epsfig {file=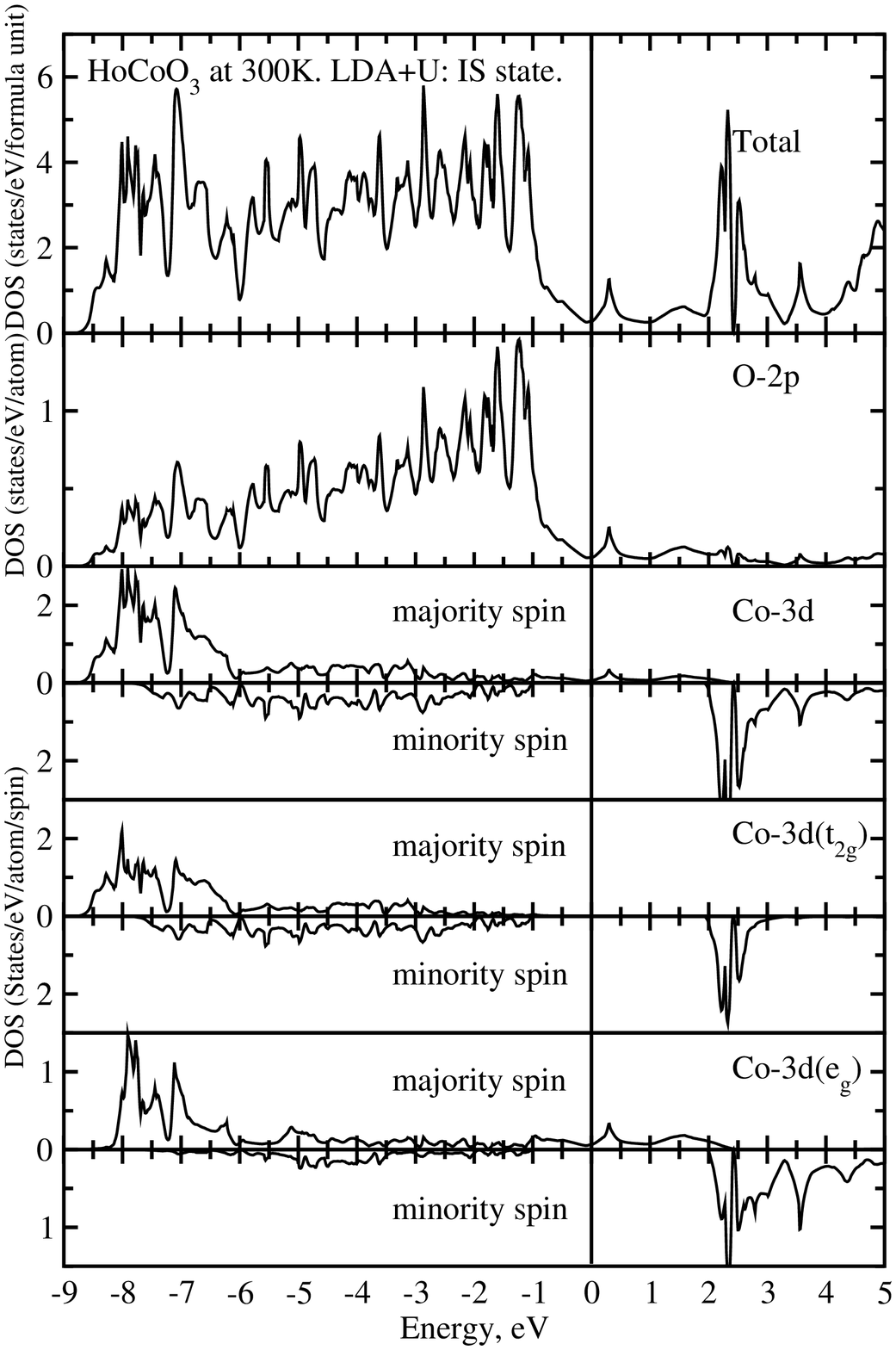,width=0.4\textwidth}
\caption {DOS of LaCoO$_3$ (left) and HoCoO$_3$ (right) calculated within the LDA+U
approach for Intermediate Spin state t$^5_{2g}$e$^1_g$ of Co ion. Top panel:
total DOS; second from top panel: partial O-2$p$ DOS; Last three panels: partial
Co-3$d$, Co-3$d$~(t$_{2g}$) and Co-3$d$~(e$_{g}$) DOS correspondingly.
The Fermi level is zero energy.}
\label {lahoco_IS}
\end {figure}

On the other hand the Intermediate Spin state solutions
(Fig.~\ref {lahoco_IS}) are very
different from LDA. At first a peak above Fermi energy
appears for minority spin Co-3$d$~(t$_{2g}$) states corresponding
to a hole in t$_{2g}$ shell and at second the majority
spin Co-3$d$~(e$_g$) band becomes half-filled (t$^5_{2g}$e$^1_g$ configuration).
The magnetic moment values on Co ions are equal to 2.2~$\mu_B$ for LaCoO$_3$
and 2.17~$\mu_B$ for HoCoO$_3$. The Intermediate State solution gave metallic state
in contrast to the semiconductor properties of LaCoO$_3$. In \cite{Korotin96}
this contradiction was explained by the possibility of orbital ordering
of the partially filled e$_g$-orbitals of Co$^{3+}$ ions in Intermediate State and opening an energy
gap in the orbitally ordered state. Recently the evidence for orbital order in
LaCoO$_3$ was found in diffraction experiments \cite{palstra}.

\begin {figure}
\vspace{3mm}
\centering
\epsfig {file=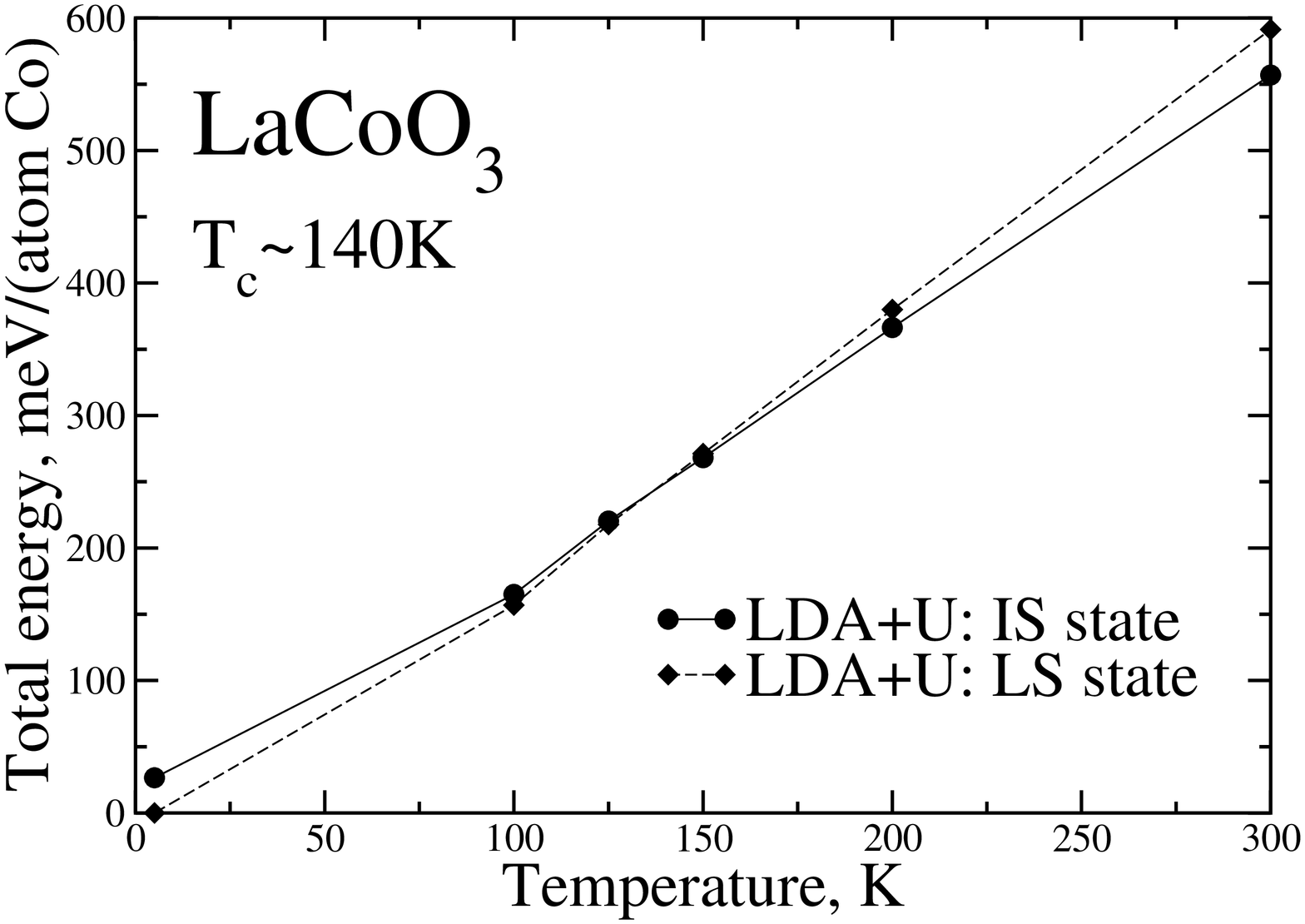,angle=270,width=0.4\textwidth}
\rule{0.05\textwidth}{0pt}
\epsfig {file=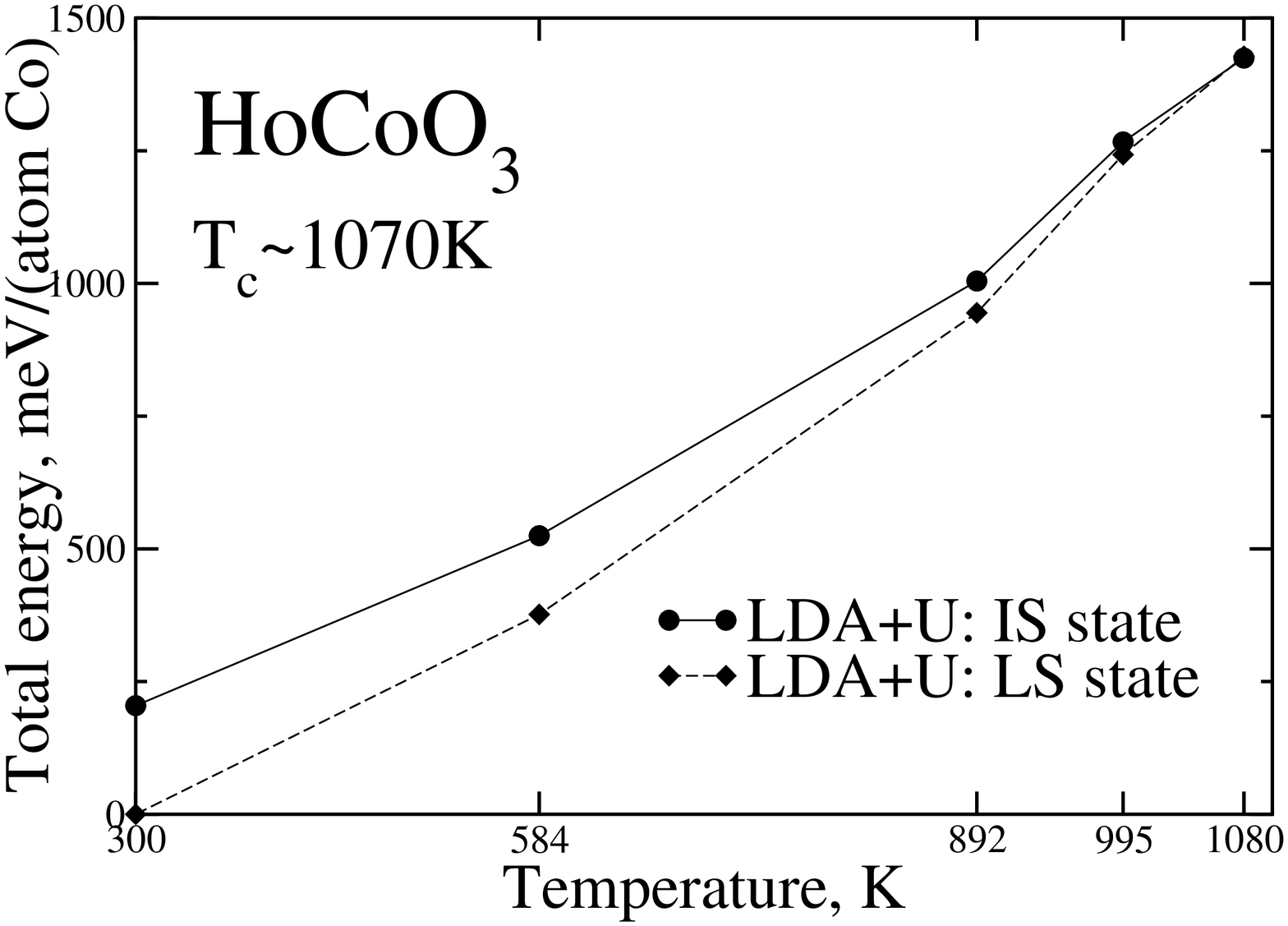,angle=270,width=0.4\textwidth}
\caption {Comparison of total energy per Co ion of Intermediate Spin state
t$^5_{2g}$e$^1_g$ (full line, circles) and Low Spin state t$^6_{2g}$e$^0_g$ (dashed
line, diamonds) solutions for LaCoO$_3$ (left) and HoCoO$_3$ (right)
calculated with the LDA+U approach as a functions of temperature.
The temperature of transition (calculated as the temperature where
two lines cross) is $\sim 140$~K for LaCoO$_3$ and $\sim 1070$~K for HoCoO$_3$.}
\label {lahoco_toten}
\end {figure}

The most interesting are results for total energy values of Low Spin and Intermediate Spin
solutions. In Fig.~\ref {lahoco_toten} those
energies are plotted as a function of the temperature (temperature
is taken into account in our calculations only via experimental crystal
structure parameters). For LaCoO$_3$ at 5~K the total energy difference between
Low Spin and Intermediate Spin states is only 37~meV and (by interpolation) becomes zero
at 140~K. For room temperature (300~K) Low Spin state is higher in energy than
Intermediate Spin on 45~meV.

For HoCoO$_3$ the situation is very different. At room temperature
there is a very large total energy difference between Low Spin and Intermediate Spin states
(200~meV) and only at 1080~K Intermediate Spin state becomes lower in energy than Low Spin.
The interpolated value of T, where energies of Low Spin and Intermediate Spin are equal is 1070~K.

One can identify the temperatures, where lines for Low Spin and Intermediate Spin states
cross in Fig.~\ref {lahoco_toten}, as magnetic
transition temperatures. The value for LaCoO$_3$ equal to 140~K agrees very well
with the 100~K maximum in experimental susceptibility curve \cite {Imada98}.
The calculated magnetic transition temperature for HoCoO$_3$ equal to 1070~K
also agrees with the results of \cite {liu},
where authors conclude that ``at room temperature all cobalt ions are
in the Low Spin state'' and only as
temperature increases to 1000~K ``the possible electronic phase transition
can be suggested''.

\section {Conclusion}

Comparative study of electronic structure and magnetic properties of LaCoO$_3$
and HoCoO$_3$ was done using LDA and LDA+U approaches. The chemical pressure
induced by  substitution of La ions by  smaller Ho ions leads to the bending of
Co-O-Co bond angles and compression of Co-O bond length that in its turn
increases the value of crystal field splitting in HoCoO$_3$. Total energy
calculations for Low Spin and Intermediate Spin states demonstrated that this
increase results in stabilization of non-magnetic solution in HoCoO$_3$ and
hence increasing of transition temperature from 140~K (LaCoO$_3$) to 1070~K
(HoCoO$_3$). Calculated transition temperatures agree well with the
experimental estimations.

\begin {acknowledgments}
The work was supported by the INTAS project No.01-0278
and the Russian Foundation for Basic Research through  grants RFFI-01-02-17063
(VA, IN, MK) and RFFI-03-02-06026,
the grant of Ural Branch of the Russian Academy of Sciences for
Young Scientists, Grant of the President of Russia for
Young Scientists MK-95.2003.02 (IN).
\end {acknowledgments}

\end {document}